**Safety and Security Dynamics in Gulf Cooperation Council (GCC) Countries:**

**A Machine Learning Approach to Forecasting Security Trends**

**Mahdi Goldani**

m.goldani@hsu.ac.ir

**Abstract:**

The GCC region — include Saudi Arabia, UAE, Bahrain, Kuwait, Qatar, and Oman— has critical geopolitical and economic importance, being rich in oil and positioned along vital maritime routes. However, the region also faces complex security challenges, ranging from traditional threats like interstate conflicts to non-traditional risks such as cyber-attacks, piracy, and environmental concerns. This study investigates the safety and security index for six GCC countries using machine learning techniques, specifically XGBoost, to forecast security trends for the next five years. Data from the Global Peace Index and World Bank development indicators were employed to construct the model. Key indicators related to economic, political, and environmental factors were selected using the Edit Distance on Real sequence (EDR) feature selection method. The model demonstrated high accuracy, with a mean absolute percentage error (MAPE) of less than 10% across all countries. The results indicate that Bahrain and Saudi Arabia are likely to experience improvements in their safety and security indexes, while Kuwait and Oman may face challenges in maintaining their current levels of security. The findings suggest that economic diversification, environmental sustainability, and social stability are critical for ensuring long-term security in the region. This study provides valuable insights for policymakers in designing proactive strategies to address emerging security threats.

## 1. Introduction

The GCC are introduced as one of the most strategically important area in the world [1]. due to its vast energy reserves, which account for a significant portion of the world's oil and natural gas production. The region's geopolitical importance is further heightened by its location at the crossroads of major shipping routes, making it critical for global trade and energy supply. Moreover, the Gulf has been a focal point for international political and military interests, with various powers vying for influence in the region. Intrastate and interstate conflicts in the GCC have been numerous. Some conflicts originated centuries ago and are still intense [2]. From the perspective of the Security, the GCC region has been a typical security subcomplex since the 1970s [3]. But, from another perspective, the strategic relevance of the GCC is also rooted in the historical and ongoing geopolitical rivalries that set the security architecture in the region. In a context where the Gulf occupies an important place in the global energy supply node, it faces quite an ambitious and intervening strategy being pursued by regional powers such as Saudi Arabia and the UAE, among others, and external actors like the United States, China, and Russia. Ideological and sectarian divisions in the region have further created an impact on its security dynamics, while the Sunni-Shia rivalry has also played a serious role in fueling not only internal conflicts but also proxy wars within the region.

The security challenges for the region are increasingly influenced by non-traditional threats such as cyber-attacks, piracy, and other environmental concerns about oil production and maritime security [4]. As such, according to Sadiq [5], balancing the various aspects of security requires a strategic synergy of military preparedness, diplomatic engagement, and regional cooperation. Nevertheless, the Gulf remains one of the most volatile and conflict-prone areas in the global political landscape.

The purpose of this study is to investigate the safety and security index for six countries in the GCC. The six countries examined in this study are Saudi Arabia, Qatar, United Arab Emirates, Kuwait, Bahrain and Oman. According to the purpose of the research, first, the most relevant variables with the safety and security index are selected from among the World Bank indices, and then by building a machine learning model and measuring the accuracy of this model, the necessary forecast for the next five years of these indices is made.

Considering the purpose of the study, the research literature is reviewed first, and then the results obtained for each country are analyzed with a review of the research methodology, and finally, the summary and conclusion of the work is presented.

## 2. literature review

The security landscape of the GCC is very complicated, with a huge number of challenges in political instability, economic transitions, environmental concerns, demographic pressures, among others. Recent studies underline how multi-faced security is, which the region needs to confront through an equally comprehensive and multi-dimensioned approach.

In the past few decades, countries in the Gulf Cooperation Council (GCC)) have made substantial economic progress due to the effective usage of massive resources of hydrocarbon. Among the above-listed GCC countries, Kuwait and UAE are recognized as the wealthiest economies across the globe based on their per capita, as oil and natural gas resources power the rapid development of the region. These stable economic conditions are one of the main factors of stability in these countries [6]. Favorable economic conditions of these countries enable these countries to increase the welfare of their citizens. Food, water, and energy security constitute a critical part of the Gulf nations' overall security framework. The countries in this region, due to overdependence on oil revenues, have sought to invest in establishing a sustainable economic model going beyond dependence on oil. The interlink between water energy and food is considered a core ingredient for attaining resource security and sustainable development in the GCC countries [7]. Siddiqui [8] shows water scarcity remains a critical challenge, with climate change intensifying the risks. Strategies such as aquaculture, integrated farming, and rainwater harvesting are proposed to address food security issues in this increasingly water-stressed environment. The security landscape is further complicated by demographic pressures-the youth bulge in the Gulf region, for instance. As Khan et al [9]. identified, an increasingly aging population demands reforms in health care and social services; if these go unfulfilled, these reforms will only serve to heighten social tensions and economic instabilities. The nature of a labor-deficient region means the government must reconsider the course of immigration policies and labor market strategy to unleash the potential of the young population. Also FDI can enhance political stability by generating economic opportunities, diverging from some existing literature that associates FDI with increased political risks[10].

Dependence on oil and the oil economy makes the countries of this region vulnerable. This dependence means that any disruption in the region, whether due to geopolitical tensions or attacks on energy infrastructure, has far-reaching consequences for the global economy and energy security [11]. Gulf countries, to overcome structural economic deficiencies, are increasingly investing their oil wealth in various sectors and acting as a driver for economic development other than oil dependent. This transformation process is considered instrumental for the long-term stability and security of their economies. The transition towards a knowledge economy and establishing free trade zones are some of the strategies that are being undertaken to build greater resilience in economic sectors and attract foreign investment [12]. The research indicates that throughout the previous several decades, continuous efforts to develop the status of ICT, education, innovation, and entrepreneurship in several GCC countries have contributed to improving their international competitiveness, as seen by advancements in rankings issued by various international organizations. Furthermore, political stability, significant financial resources, and a stable credit rating provide these countries with solid foundations for future sustainable development. However, several functional, structural, and cultural factors challenge the diversification process and the shift toward a knowledge-based economy [13].

The geopolitics of the GCC is constituted by the interplay of influence wrought by extra-regional powers and regionally operational rivalries. As Jain and Oommen [14] have mentioned, such a hegemonic stability situation does indeed imply the critical role played by external actors in shaping the security dynamics of this region. These apprehensions underpin a lot of foreign policy decisions made by the Gulf states. More disturbingly, sectarian strife that contributes to this increasingly bad security environment is fostered by outside powers [15]. The GCC remains crucial for global energy markets, with more than 80% of crude oil and condensates passing through the Strait of Hormuz destined for Asian markets such as China, India, Japan, and South Korea. This dependence means that any disruption in the region, whether due to geopolitical tensions or attacks on energy infrastructure, has far-reaching consequences for the global economy and energy security [16].

Notwithstanding the large volume of research on different aspects of GCC security, some significant knowledge gaps persist. There is a particular need for further empirical research into the interactions between demographic change and economic strategies, especially in respect to healthcare and social services, and for a more-detailed investigation of the implications of climate change for resource security. Future research should also investigate the impact of external powers on regional stability, including how external involvement shapes intra-regional relations and security strategies.

Security in the GCC states is multivariate, and hence an approach in which environmental, economic, and demographic factors are concerned is required. Their stability and long-term security will, however, be highly dependent on solving these issues with cooperation as the region continues to evolve. Further research in the identified knowledge gaps will build our comprehension of the complex dynamics at play and inform effective policy responses.

## 3. Dataset and methodology

### 3-1. Dataset

The safety and security index are targeting value of this study which is a subset of the global peace index. This index is produced by the international think-tank the Institute for Economics & Peace (IEP). The GPI report presents the most comprehensive data-driven analysis to date on peace, its economic value, trends, and how to develop peaceful societies. The report covers 99.7% of the world's population and uses 23 qualitative and quantitative indicators from highly respected sources to compile the Index. These indicators are grouped into three key domains: Ongoing Conflict, Safety and Security, and Militarization. The investigated region of this study is the GCC region, and six countries of this region are investigated. Six target countries are Saudi Arabia, United Arab Emirate, Bahrain, Kuwait, Qatar and Oman. The safety and security index of this country are extracted from international think-tank the Institute for Economics & Peace reports.

In every machine learning forecasting model opting for the appropriate predictors is vital to gain an accurate model. Due to the predictors help the model training process. Most researchers use literature to choose predictors. This way of selecting variables has its advantages and disadvantages. To build an accurate prediction model, it is necessary to choose comprehensive predictors. Consequently, this study used World Bank development indicators to opt for the appropriate predictors for each country. The World bank reports 266 indicators for 217 countries and 49 regions in each year. Indicators illuminate the economic, social, political and environmental circumstance in each country and area therefore, it is considered as a comprehensive dataset. In this study, all indicators of the World Bank are used to find out which indicators resemble political stability and are good explainers of target value. Of the 266 indicators, 20s are chosen by EDR feature selection methods. Based on Goldani and Asadi [17], The Edit Distance on Real sequence (EDR) method was chosen for feature selection due to its lower sensitivity to sample size and simpler calculations.

### 3-2. methodology

Machine learning (ML) is a powerful tool for making accurate and reliable predictions and, a sub-category of computational intelligence techniques mainly employed for deriving definitive information out of large sets of data for pattern recognition, classification, function approximation, and so forth. With the availability of vast datasets in the era of Big Data, producing reliable and robust forecasts is of great importance [18, 19, 20]. XGBoost is used in this study as an ensemble learning-based algorithm, where a set of base models are combined to create a model that obtains better performance than a single model [21]. It is considered a good method due to its combination of accuracy, efficiency, and flexibility. It surpasses in handling large datasets, managing missing data, preventing overfitting, and offering various tuning options. Its ability to process data quickly, handle imbalanced datasets, and provide interpretable models makes it a go-to choice for many machine learning practitioners. Whether for academic research or industrial applications, XGBoost remains a top choice for building robust and effective predictive models.

Table1. Comparison of XGBoost with Other Prediction Methods in Machine Learning

| Method | Strengths | Weaknesses | Best Use Cases |
|---|---|---|---|
| XGBoost | High accuracy, speed, handles large datasets, built-in regularization to prevent overfitting | More complex to tune, computationally intensive for large hyperparameter grids | Large datasets, competitions, handling missing data, high performance required |
| Random Forest | Good for preventing overfitting, easy to interpret, works well with categorical data | Less efficient on very large datasets, less accurate than boosting algorithms | Smaller or medium-sized datasets, when interpretability is important |
| Support Vector Machines (SVM) | Effective for complex decision boundaries, works well on smaller datasets | Computationally expensive for large datasets, sensitive to feature scaling | Smaller datasets, complex classification problems, high-dimensional spaces |
| Neural Networks | Excels with unstructured data, can model complex relationships | Prone to overfitting, requires large datasets, complex to tune | Unstructured data (e.g., images, text), when deep feature extraction is needed |

Source: [22]

### 3-2-1. Extreme gradient boosting (XGBoost)

Chen and Guestrin [23] have created Extreme gradient boosting (XGBoost). XGBoost stands for "Extreme Gradient Boosting", where the term "Gradient Boosting" originates from the paper Greedy Function Approximation: A Gradient Boosting Machine, by Friedman [24]. XGBoost is based on the gradient boosting algorithm, a key method in ensemble learning. It combines weak classifiers to create a stronger model, enhancing efficiency and flexibility compared to a single model. By iteratively building decision trees, XGBoost improves classification performance.

A salient characteristic of objective functions is that they consist of two parts: training loss and regularization term Eq. (1)

$$obj^{(t)} = l(f_t) + \Omega(f_t) \tag{1}$$

In Eq. (1), $f_t$ representing the t-th tree model, $l(f_t)$ is the loss function in the risk prediction, $\Omega(f_t)$is the regular term used to reduce overfitting, which can be expressed as Eq.(2)

$$\Omega(f_t) = \gamma T + \frac{1}{2}\lambda\|\omega\|^2 \tag{2}$$

In Eq. (2), T represents the number of leaf nodes in the t-th decision tree, $\gamma$ and $\lambda$ can decide penalty strength together, $\omega$ representing the weight value on each leaf node. The training loss measures how predictive model is with respect to the training data. A common choice of $L$ is the mean squared error, which is given by Eq.(3)

$$l(f_t) = \sum_i (y_i^t - \hat{y}_i^t)$$
(3)

The prediction results of the model are the weighted sum of all the decision trees, when the t-th iteration is performed, the prediction result can be expressed by Eq(4).

$$\hat{y}_i^{(t)} = \sum_{k=1}^{k} f_{(x_i)=}\, \hat{y}_i^{(t-1)} + f_t(x_i),\ f_k \epsilon F \tag{4}$$

In Eq. (4), $f_t(x_i)$ represents the t-th tree model, F is the decision tree space, is also the set of all sample risk prediction decision trees. Here $\hat{y}_i^{(t)}$ represents the prediction results of the sample I after the t-th iteration, and $\hat{y}_i^{(t-1)}$ represents the prediction results of the previous t-1 trees. Therefore, the objective function can be expressed as the formula for Eq. (5):

$$obj^{(t)} = \sum_{i=1}^{n} l\left(y_i, \hat{y}_i^{(t)}\right) + \Omega(f_t) + costant \quad (5)$$

Over fitting is common and undesirable machine learning methods behavior. Overfitting is the production of an analysis that corresponds too closely or exactly to a particular set of data and may therefore fail to fit additional data or predict future observations reliably. One of the solutions to avoid over fitting is tuning parameters to maximize model performance. There are a bunch of hyperparameter tuning methods. By considering the small dataset, Grid Search is a great option because it exhaustively explores all combinations of hyperparameters. The list of hyperparameters of XGboost is represented in table2.

Table2. Hyperparameters of XGboost

| HYPERPARAMETER | DESCRIPTION | TYPICAL VALUES |
|---|---|---|
| **N_ESTIMATORS** | Number of trees (or boosting rounds). | 100, 500, 1000 |
| **LEARNING_RATE** | Step size shrinkage used to prevent overfitting. | 0.01, 0.1, 0.3 |
| **MAX_DEPTH** | Maximum depth of each decision tree. | 3, 5, 7, 10 |
| **MIN_CHILD_WEIGHT** | Minimum sum of instance weight (Hessian) needed in a child node. | 1, 3, 5 |
| **GAMMA** | Minimum loss reduction required to make a further partition on a leaf node. | 0, 0.1, 0.5, 1 |
| **SUBSAMPLE** | Fraction of samples used for building each tree. | 0.6, 0.8, 1 |
| **COLSAMPLE_BYTREE** | Fraction of features to be used for building each tree. | 0.6, 0.8, 1 |
| **COLSAMPLE_BYLEVEL** | Fraction of features used per tree at each level of boosting. | 0.6, 0.8, 1 |
| **LAMBDA (L2 REGULARIZATION)** | L2 regularization term to control model complexity and prevent overfitting. | 0, 1, 5, 10 |
| **ALPHA (L1 REGULARIZATION)** | L1 regularization term to control model complexity and sparsity. | 0, 1, 5, 10 |
| **SCALE_POS_WEIGHT** | Balancing of positive and negative weights for imbalanced classification. | 1, depending on class imbalance |

### 3-2-2. Model Evaluation

Due to the limited sample size, the cross-validation method is chosen to obtain reliable insights. Cross-validation is particularly useful for small datasets because it maximizes the use of available data by repeatedly dividing the dataset into training and validation sets.

The mean absolute percentage error (MAPE) is employed to assess the prediction accuracy for both training and testing data. MAPE is calculated as follows:

$$\text{MAPE} = \frac{1}{n}\sum_{i=1}^{n}\left|\frac{A_i - F_i}{A_i}\right| \times 100 \quad (6)$$

where $A_i$ is the actual value and $F_i$ is the forecasted value. This metric provides a straightforward and interpretable measure of forecast accuracy [25].

## 4. Result

In prediction models, predictors help models to have more accurate forecasting. Ten predictors for six countries (dataset) were selected using the EDR method. Fig 2 shows how much each variable affects the target variable. Looking at selected features in each dataset, most selected features are economic. It admits that security and economic strength are inextricably linked. There can be no lasting peace – and therefore no collective security – without sustainable progress towards prosperity. GNI, foreign investigation, and growth in GDP are the most common indicators that are linked to security.

Fig1. Feature importance

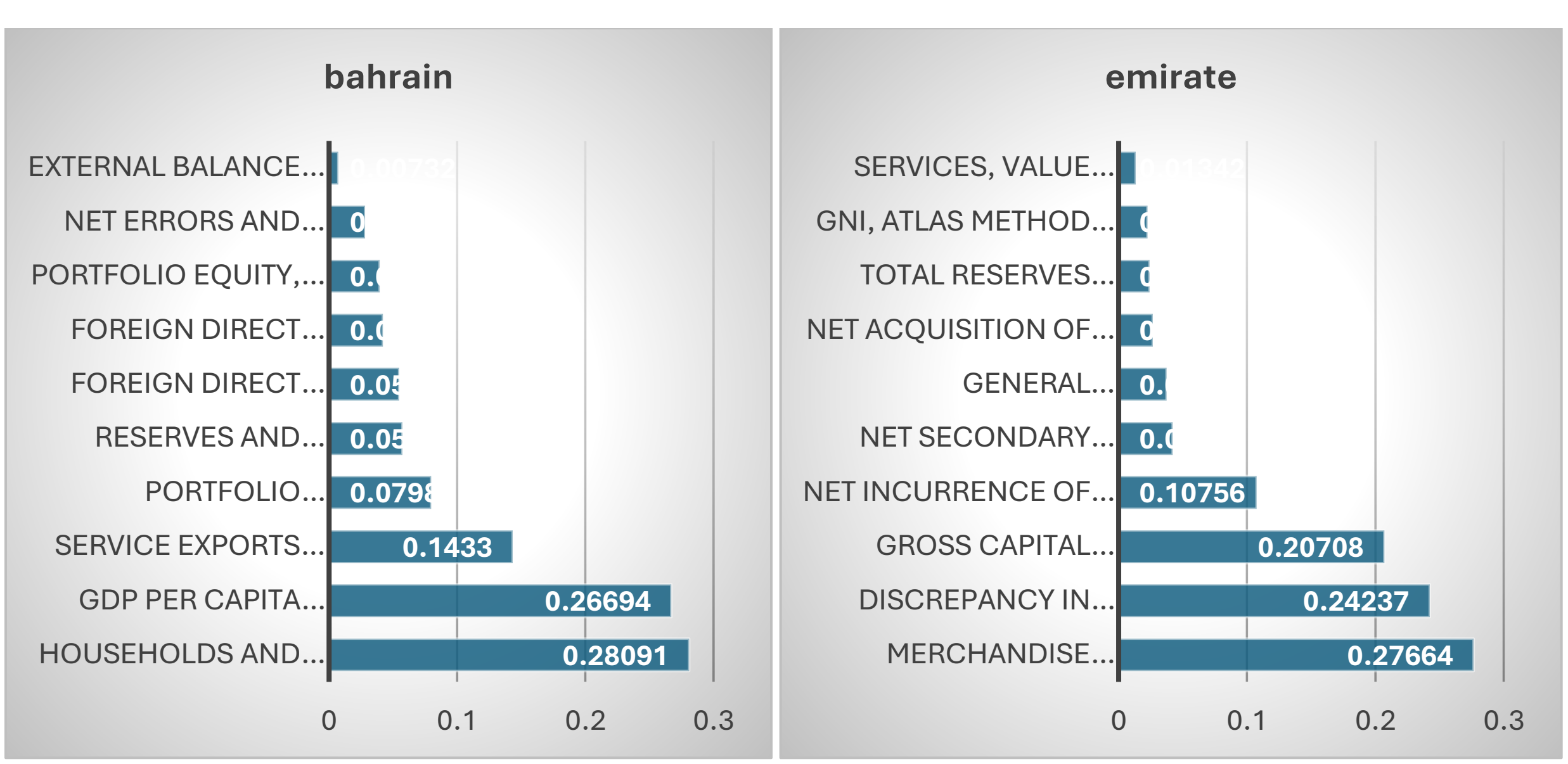

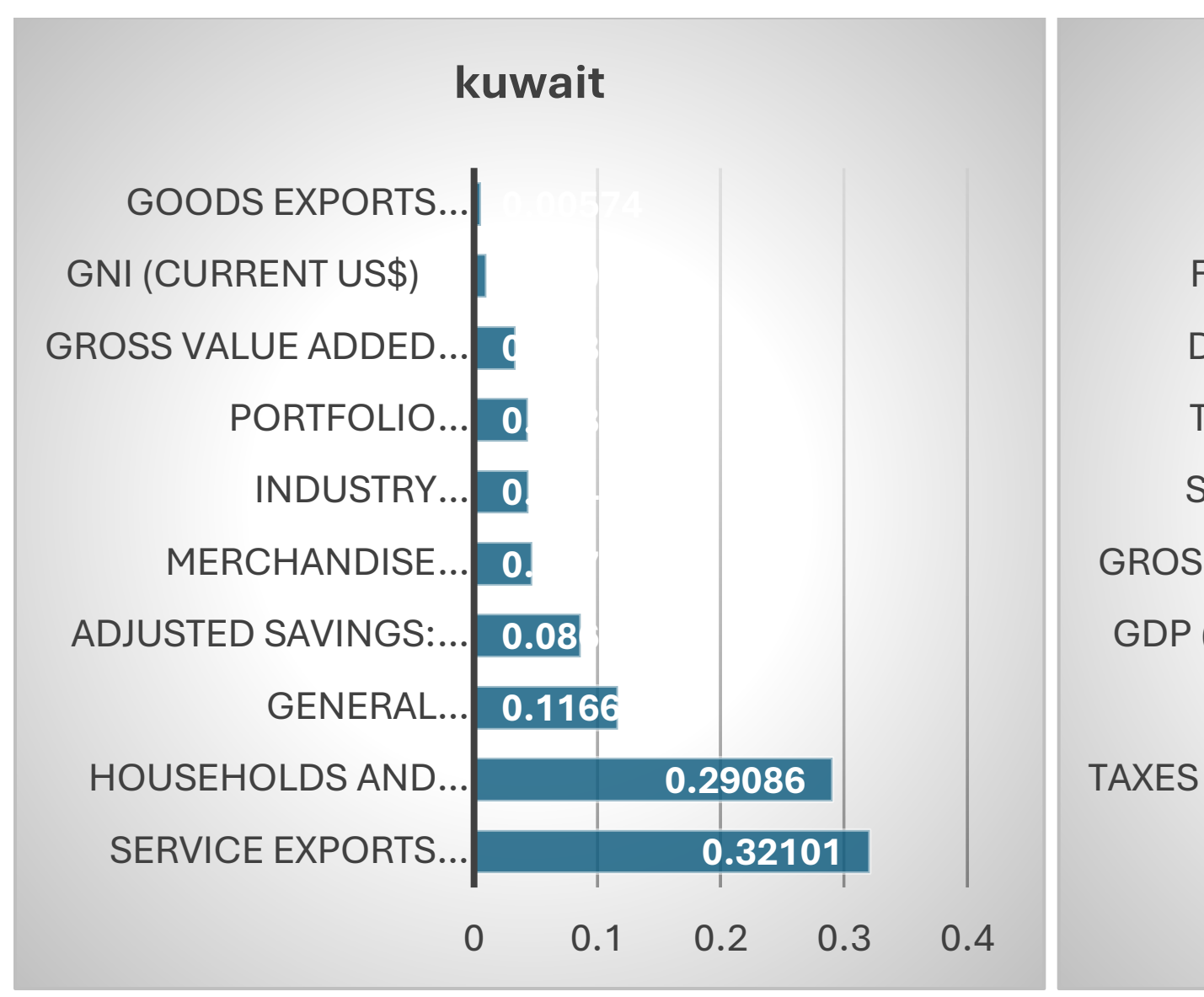

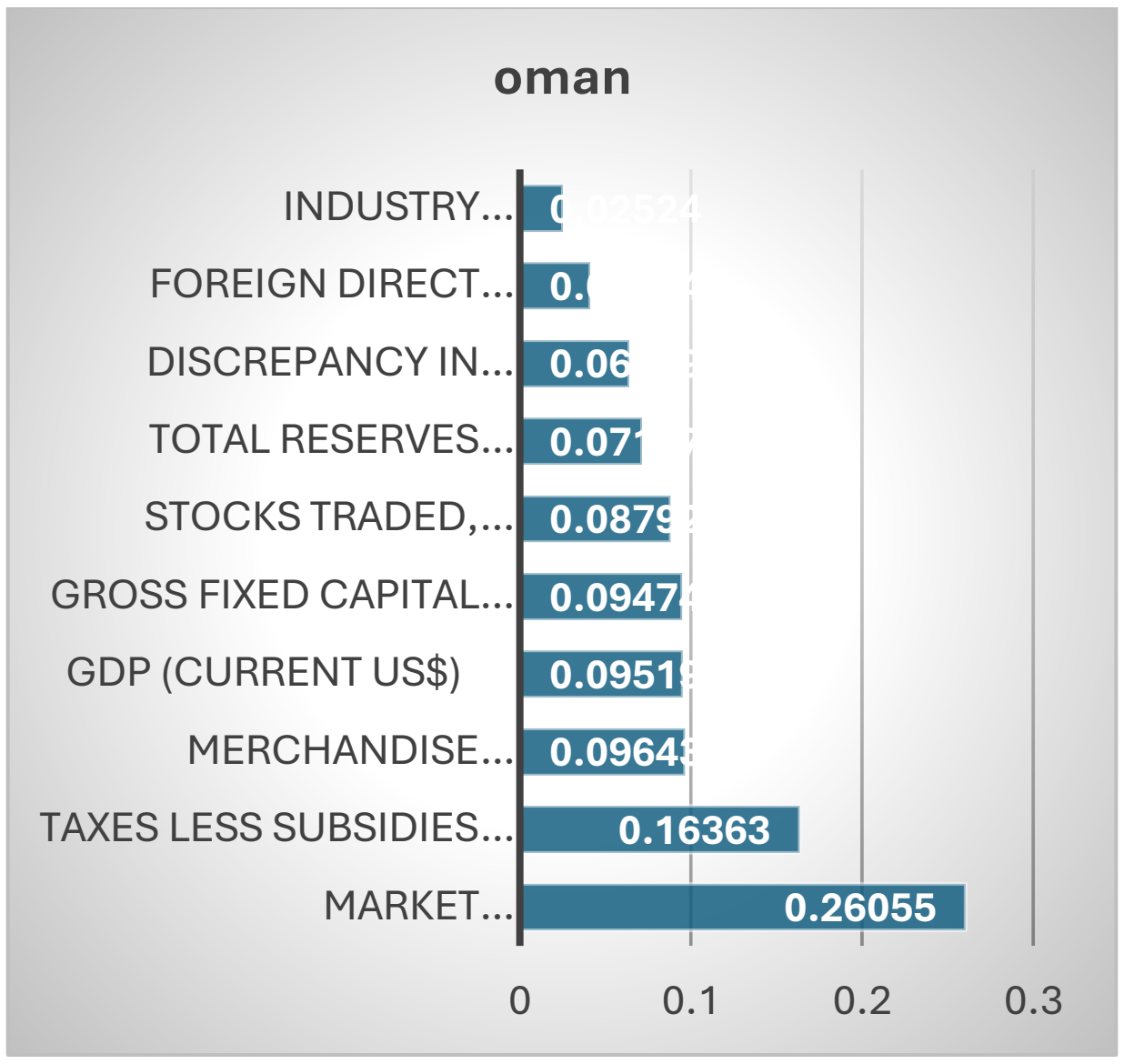

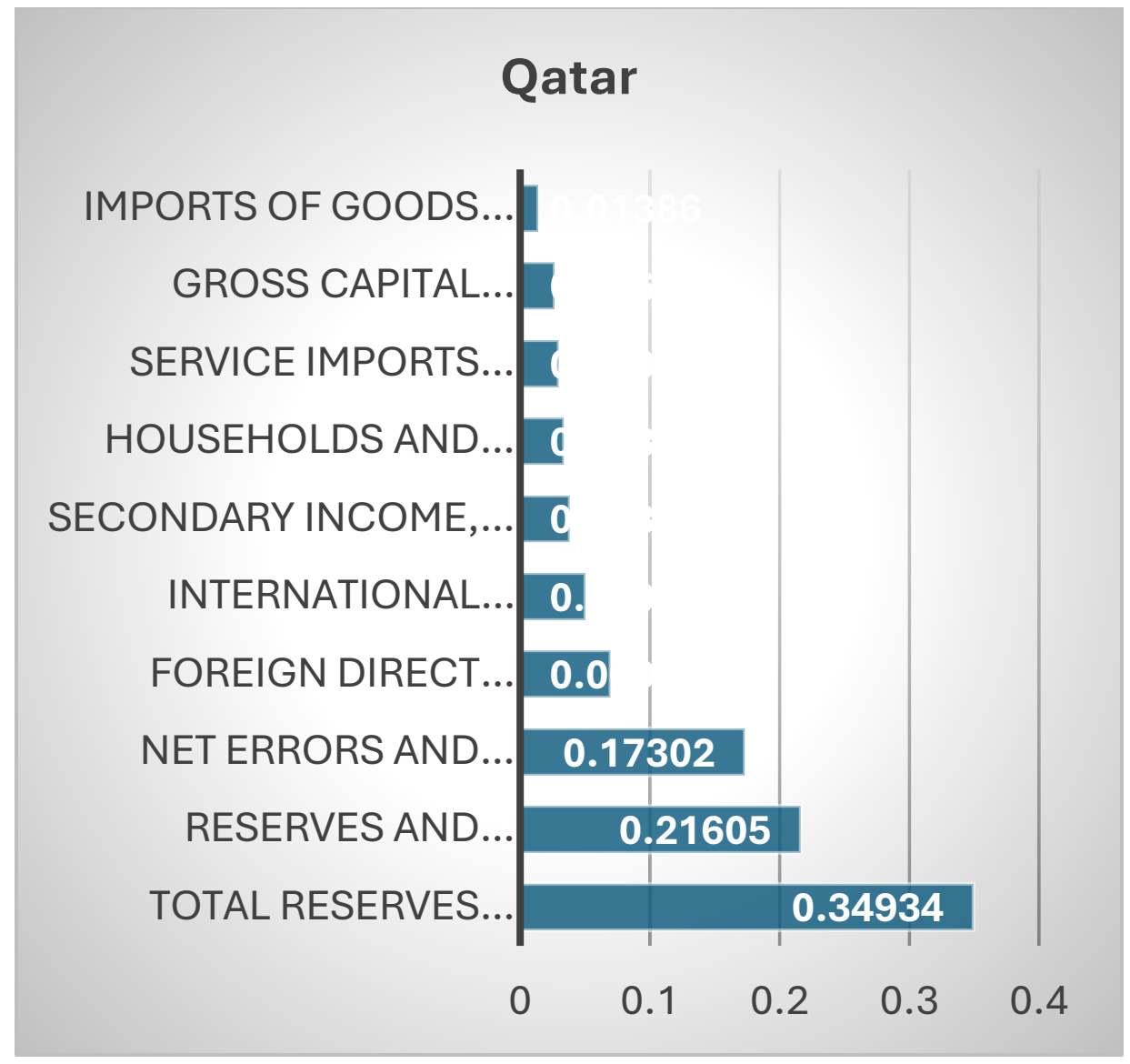

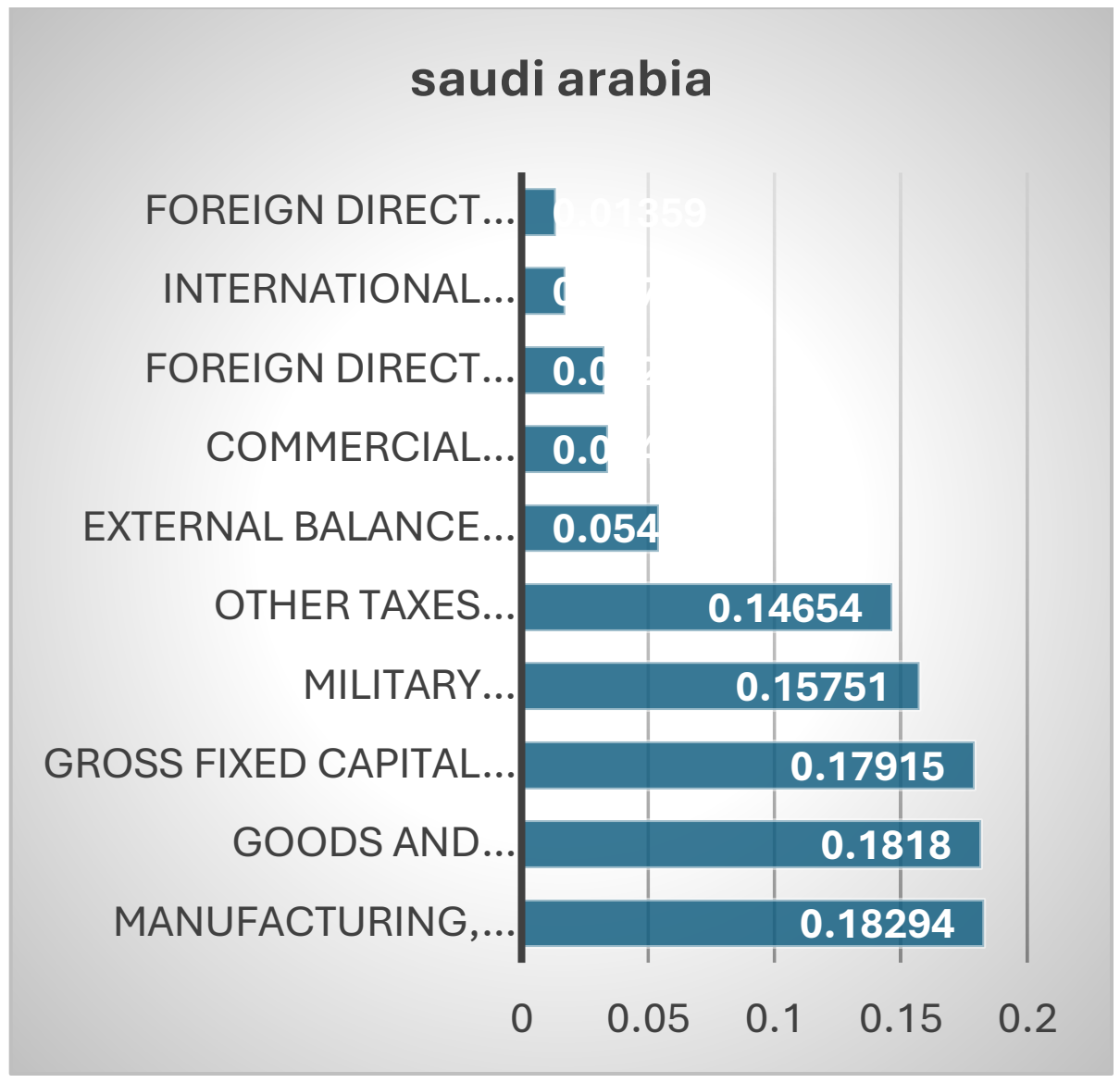

In this paper, the optimal parameters were determined by grid search and cross-validation, specific parameters that can improve the model accuracy. The data from 2008 to 2017 were used to train the model and the remaining data from 2017 to 2023 were considered as test data. Table 3 shows the accuracy in predicting training data and test data in the six countries studied. The value of MAPE corresponding to all six countries is less than 10%, which indicates the high accuracy of prediction of both data set and training. Among the available datasets, Oman and the United Arab Emirates have the highest test data prediction accuracy.

Table3. MAPE OF Training And Testing

| MAPE | Train (2008-2017) | Test (2018-2023) |
|---|---|---|
| bahrain | 2.168993 | 2.502727 |

| emirate | 0.022686 | 2.391801 |
| --- | --- | --- |
| kuwait | 0.021452 | 3.817345 |
| oman | 2.903549 | 2.235817 |
| qatar | 0.025653 | 3.083447 |
| saudi arabia | 0.019528 | 3.916139 |

Fig2. MAPE OF Training And Testing

To find a complete view of the accuracy of predicting the safety and security index in each country, the Fig (2) is presented. The graph shows that the XGBoost method has better predicted the changes in the countries of Saudi Arabia, Bahrain and Qatar. Among the existing countries, the country of Emirate has an upward trend in the safety and security index. The forecast is almost bullish. The trend of the safety and security index in the remaining 5 countries is downward. A look at the forecasts made for these five countries shows that, except for Oman, the rest of the trends are downward and correctly predicted.

Fig3. Actual vs Predicted safety and security index for GCC Nations

Bahrain

United Arab emirate

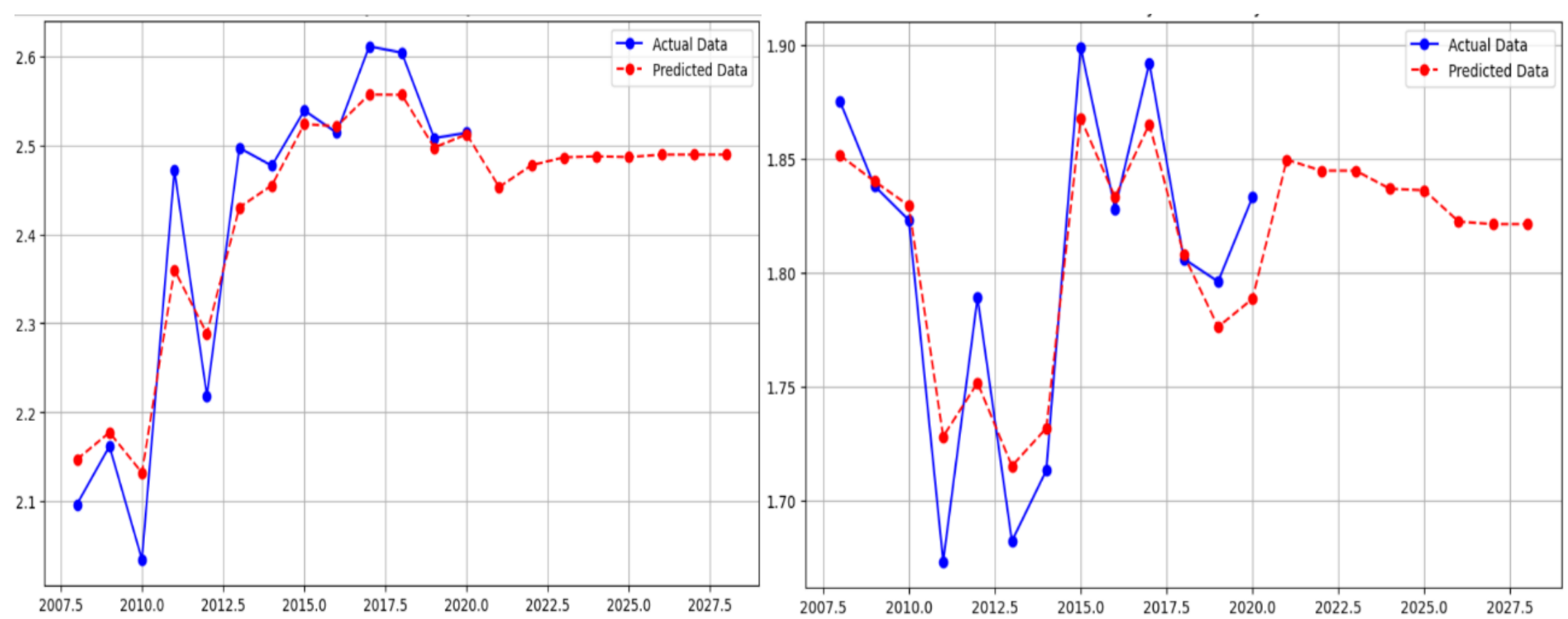

Kuwait Oman

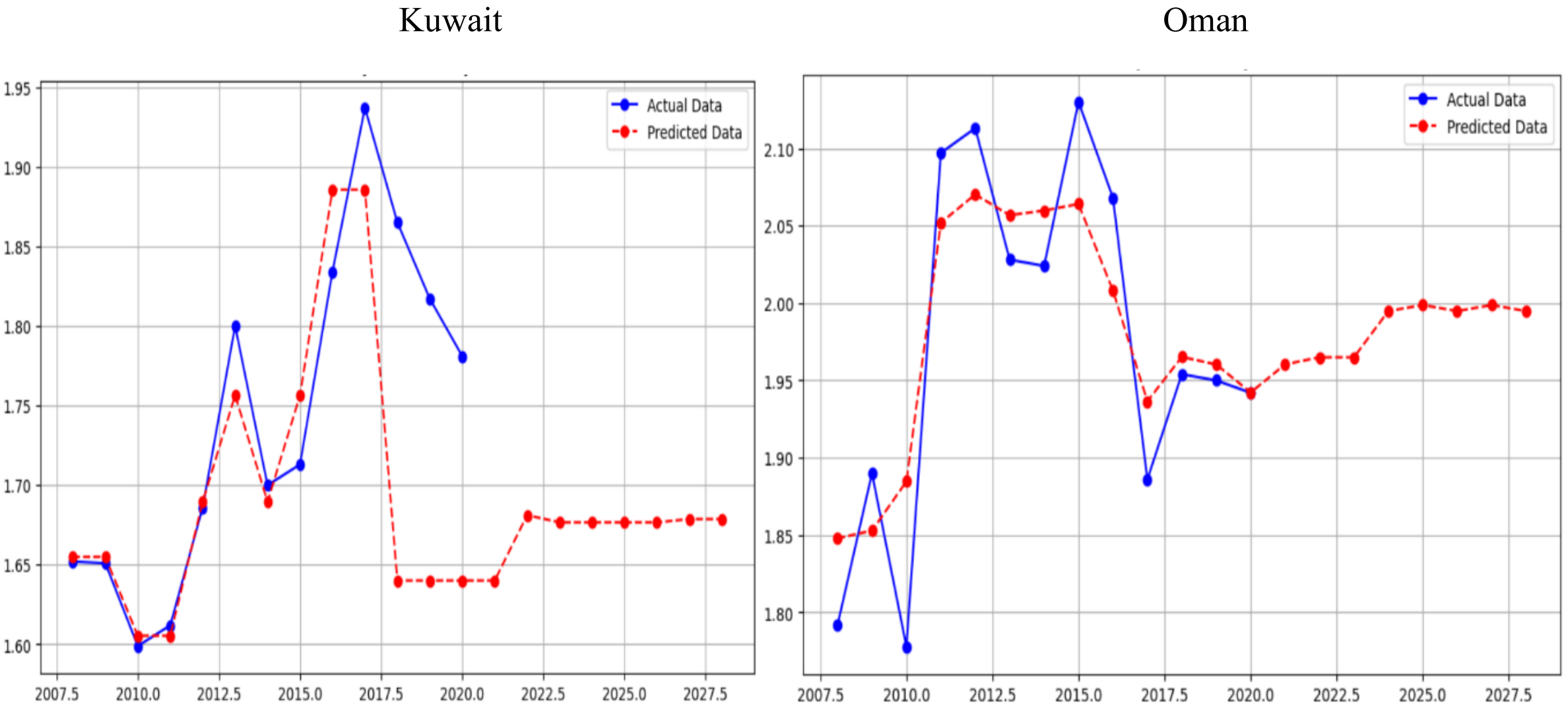

Qatar Saudi Arabia

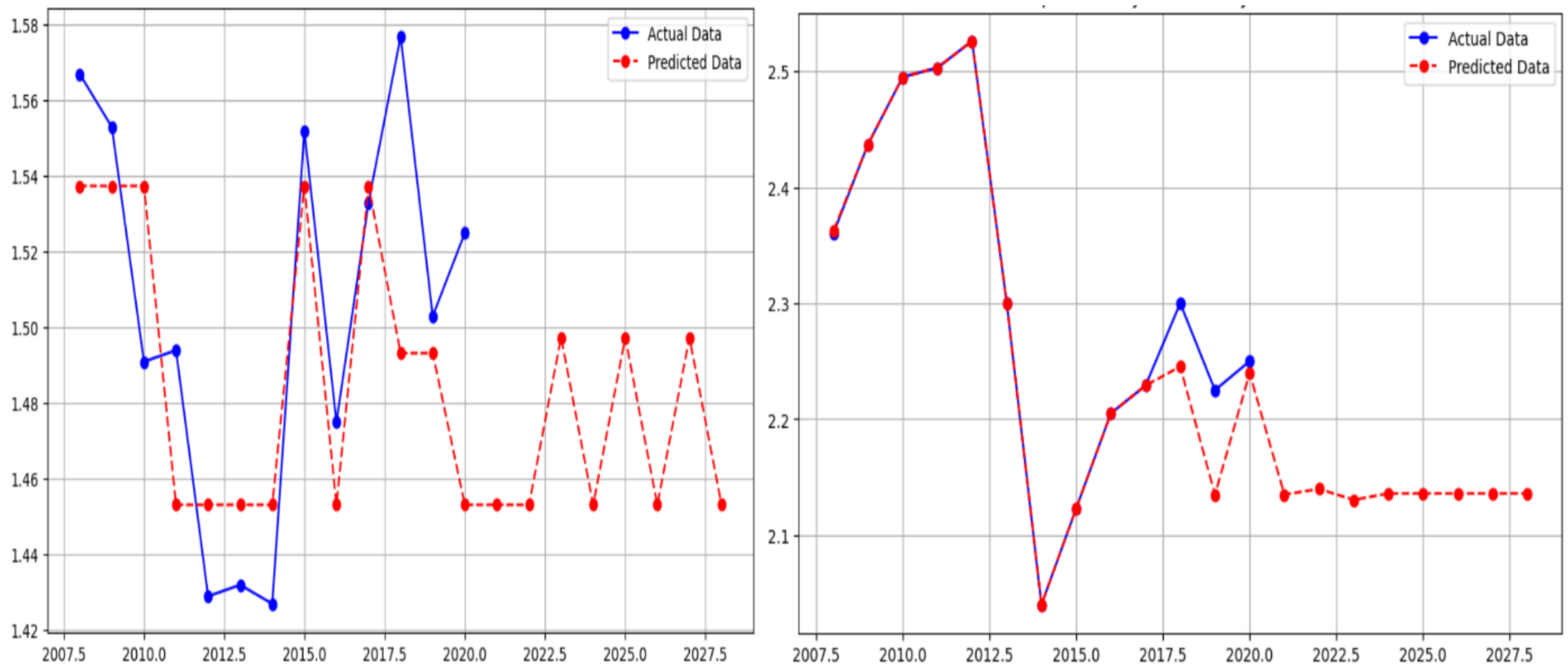

After the training to build the model and testing to obtain the accuracy of the model prediction, the predictor variables are simulated and then based on these predictors, the safety and security index values are predicted for the next 5 years.

**Bahrain:** Based on the results, Bahrain improved somewhat through 2018 regarding the categories of safety and security, or those things influencing it. Since then, things may have seen slight challenges. The model now forecasts that the conditions will flatten out and remain relatively flat going forward, perhaps to reflect an assumption of continued moderate growth and political stability, although without large fluctuations in safety and security. The features in the dataset, which are most economical, such as GDP per capita and investment levels, apparently influence the model of prediction. These indicators of safety and security in Bahrain will show the same trend in world country stability.

**United Arab Emirate:** This trend analysis may lead to the inference that the safety and security of the UAE are directly proportional to its economic conditions, namely merchandise trade, and investment, Gross Capital Formation; being heavily dependent on the same, any disturbance in either of these can affect stability directly. It foresees a future relatively stable for the UAE post-2020, probably reflecting expected recovery and stability in trade and government spending.  In other words, the UAE economic indicators are strong yet relatively sensitive; safety and security there are partly based on economic stability, particularly trade and governmental expenditures. The forecasts show cautious optimism on the ground that the economy and security situation will be relatively stable for some time in the future.

**Kuwait:** As observed in the safety and security forecast for Kuwait, service exports and household consumption levels set the pace, economic diversification into non-oil sectors is the clue to stability. Government spending is important in sustaining stability, especially as it pertains to public services and infrastructure. This trend suggests that, from 2010 to 2017, Kuwait was going through a period of growth and stability but started declining, possibly due to economic challenges or other geopolitical events. Looking into the prediction model, one should be expecting a future where it is going to remain stable with a low level of safety and security, reflecting the idea that the country might not experience the same level of growth in the same manner, at least according to the features used in this model. This steady predicted trend post-2020 reflects cautious optimism, entailing the assumption that salient features of Kuwait's economy will remain stable without a significant rise in safety/security-related outcomes. Thus, given the high value of service exports, further diversification away from oil and into other sectors will be pivotal in ensuring Kuwait's stability going forward.

**Oman:** The major determinants of safety and security prediction in Oman are its market capitalization and fiscal policies, inclusive of taxes and subsidies. In such a country, the position of financial markets is central in predicting stability, which is appropriately reflected in the key position of market capitalization, stocks traded, and foreign investment. A stable, growing economy, particularly through investments in infrastructure and the confidence of foreign investors, can help improve internal security and public services further in Oman. The fig3 suggests that in recent years, Oman has experienced some fits and starts, such as in 2014-2017, doubtless connected with exogenous shocks like the oil price and regional turbulence. However, the prediction model assumes a gradual upward improvement and stabilization after 2020, presumably reflecting assumptions of economic recovery, strengthened reserves, and sustained investment in infrastructure as well as capital formation. The post-2020 steady predictions illustrate expectations of moderate but consistent improvements in the safety and security-related indicators.

**Qatar**: The forecast of safety and security in Qatar is closely related to its financial strength, represented mainly by the total reserves and related economic items. From these features, it is clear that the wealth of Qatar is essentially important in maintaining stability and in taming both internal and external risks. Leveraging this strong reserve base allows Qatar to keep public spending robust, supporting its security apparatus even during periods of economic turmoil. The graph below captures the strong performance of Qatar historically but also a high degree of volatility, most likely informed by exogenous factors such as oil prices and regional dynamics. The forecast for safety and security is more stable, if lower, into the future, which would suggest that one expectation is for Qatar to continue in a state of moderate security without many of the changes from one year to another that has characterized past performance.

**Saudi Arabia:** Saudi Arabia's safety and security are thus very deeply intertwined with its manufacturing sector and export revenues, especially regarding oil. Another major factor contributing to the stability of the country includes infrastructure investments and military spending. As can be seen from the forecasted model, although the country has experienced volatility in the past, it is expected to have relative stability in the forward-going future, probably due to sustained government spending and a preoccupation with the balance of trade and foreign investment. The forecasted graph indicates that even though there had been quite a flux previously, from 2011 to 2017 specifically, the model estimates Saudi Arabia to be much safer and more secure beyond 2020. This simply reflects the model assumptions of the critical economic indicators that include manufacturing, exports, and government spending, remaining stable for a safe and stable environment in the forward-going years.

## 5. Conclusion

The GCC region is a strategic pivot in the world, given the immense deposits of oil and natural gas, and its privileged situation in world maritime routes. The security of the region is equally important, as it continues to be a focal point of international interest and rivalry, which has evolved into full-scale conflicts. As demonstrated in this study, the security landscape in the GCC is multilayered, multidimensional, and defined by both traditional and non-traditional threats. The study analyzed the safety and security index for six GCC countries, namely Saudi Arabia, United Arab Emirates, Bahrain, Kuwait, Qatar, and Oman.

### Economic and security

The literature shows that economic diversification, geopolitical dynamics, environmental challenges, and demographic pressures have combined to shape the security landscape of the GCC. The results of identifying the dimensions that affect the security and stability of the region show that the economic element is still the main axis in determining stability within the region. This finding has been confirmed in accordance with the works of Almasri [6], Ulrichsen [11], Ewers [12], and Ben Hassen [13]. The GCC countries are extremely dependent on their oil and gas exports. Economic stability is, accordingly, very sensitive to changes in world oil prices. The government budgets suffer when these prices have been substantially reduced, which leads to a decrease in public expenditure and, accordingly, to a lower level of living-which might increase social unrest. Most Gulf countries have some very generous social benefits: extensive subsidies, free healthcare, and education, all paid for by oil revenues. These policies are very important for maintaining popular support, and any economic stringency cutbacks could give rise to social tension. Hence, the countries of the Gulf try to attract FDI for the purpose of economic growth and diversification. A favorable investment climate requires political stability and security. An ailing economy or one with risk scaring investors will have implications for economic development and perceptions of security.

In conclusion, economic diversification, fiscal resilience, and social welfare optimization are essential to securing the stability of GCC countries in the face of oil price volatility and shifting global economic dynamics. Accelerating the growth of non-oil sectors, supporting SMEs, and incentivizing FDI in key areas can reduce dependency on oil, while targeted social programs and job creation initiatives, especially for youth and women, can sustain public support and mitigate potential unrest. Strengthening fiscal policy through countercyclical spending and effective management of sovereign wealth funds will create essential buffers for economic downturns, enabling continued public investment in critical areas without compromising stability. A favorable investment environment, supported by transparent governance, legal protections, and public-private partnerships, can further attract foreign capital essential for diversified growth and long-term security.

Additionally, sustainable environmental policies and strong regional cooperation are crucial to securing the region's future stability. Investing in renewable energy, water security, and climate resilience aligns the GCC economies with global sustainability goals and reduces long-term environmental vulnerabilities. Enhanced regional collaboration on economic, security, and counter-terrorism efforts can further strengthen resilience, while robust digital infrastructure and cybersecurity initiatives support economic modernization in an increasingly digital world. Together, these measures provide a framework for GCC countries to maintain security and stability through dynamic economic and social strategies, laying the foundation for a resilient, diversified, and prosperous future.

**Reference**

1. Potter LG, Sick GG, editors. *Security in the Persian Gulf*. New York: Palgrave Macmillan; 2002.
2. Askari H. Catalogue of Persian Gulf Conflicts. In: *Conflicts in the Persian Gulf: Origins and Evolution*. New York: Palgrave Macmillan US; 2013. p. 1-30.
3. Han J, Hakimian H. The regional security complex in the Persian Gulf: The Contours of Iran's GCC Policy. *Asian J Middle East Islam Stud*. 2019;13(4):493-508.
4. Goldani M. A comparative comparison of the most important variables of behavior with the political stability index, between MENA countries and OECD member countries. *Middle East Stud*. 2024;30(4):59-81.
5. Sadiq K. *The Arabian Gulf and the World Economy*. Cambridge: MIT Press; 2017.

6. Almasri RA, Narayan S. A recent review of energy efficiency and renewable energy in the Gulf Cooperation Council (GCC) region. *Int J Green Energy*. 2021 Nov 14;18(14):1441-68.
7. Ulrichsen KC. The Geopolitics of Insecurity in the Horn of Africa and the Arabian Peninsula. *Middle East Policy*. 2011;18(2).
8. Siddiqui S. Ecosystem Services, Climate Change, and Food Security. In: *Handbook of Research on Sustainable Development Goals, Climate Change, and Digitalization*. IGI Global; 2022.
9. Khan H, Hussein S, Deane J. Nexus Between Demographic Change and Elderly Care Need in the Gulf Cooperation Council (GCC) Countries: Some Policy Implications. *Ageing Int*. 2017;42:466-87.
10. Okara A. Does foreign direct investment promote political stability? Evidence from developing economies. Economic Modelling. 2023 Jun 1;123:106249.
11. Ulrichsen KC. Rebalancing regional security in the Persian Gulf. *Rice University's Baker Institute for Public Policy, Center for the Middle East*. 2020;14-5.
12. Ewers MC, Malecki EJ. Leapfrogging into the knowledge economy: Assessing the economic development strategies of the Arab Gulf States. *Tijdschr Econ Soc Geogr*. 2010;101(5):494-508.
13. Ben Hassen T. The GCC Economies in the Wake of COVID-19: Toward Post-Oil Sustainable Knowledge-Based Economies? *Sustainability*. 2022;14(18):11251.
14. Oommen GZ. South Asian migration to the GCC countries. In: *South Asian Migration to Gulf Countries: History, Policies, Development*. 2017;17:37-65.
15. Pasha AK. Political turbulence in West Asia: impact on India. In: *India's National Security*. Routledge India; 2016. p. 278-88.
16. Ulrichsen KC. Rebalancing regional security in the Persian Gulf. *Rice University's Baker Institute for Public Policy, Center for the Middle East*. 2020 Feb;14-5.
17. Goldani M, Asadi Tirvan S. Sensitivity Assessing to Data Volume for forecasting: introducing similarity methods as a suitable one in Feature selection methods. *arXiv e-prints*. 2024.
18. Campisi G, Muzzioli S, De Baets B. A comparison of machine learning methods for predicting the direction of the US stock market on the basis of volatility indices. *Int J Forecast*. 2024;40(3):869-80.
19. Kargah-Ostadi N. Comparison of machine learning techniques for developing performance prediction models. In: *Computing in Civil and Building Engineering (2014)*. 2014; p. 1222-9.
20. Masini RP, Medeiros MC, Mendes EF. Machine learning advances for time series forecasting. *J Econ Surv*. 2023;37(1):76-111.
21. Chen Z, Fan W. A freeway travel time prediction method based on an XGBoost model. *Sustainability*. 2021;13(15):8577.
22. Goldani M. Comparative analysis of missing values imputation methods: A case study in financial series (S&P500 and Bitcoin value data sets). *Iranian J Finance*. 2024;8(1):47-70.
23. Chen T, Guestrin C. XGBoost: a scalable tree boosting system. In: *Proceedings of the 22nd ACM SIGKDD International Conference on Knowledge Discovery and Data Mining*. 2016 Aug.
24. Friedman JH. Greedy function approximation: a gradient boosting machine. *Ann Stat*. 2001;29(5):1189-232.

25. Hyndman RJ, Koehler AB. Another look at measures of forecast accuracy. *Int J Forecast*. 2006;22(4):679-88.